\documentclass[12pt]{article} \textwidth=6in \oddsidemargin=0in
\usepackage{amssymb} 
\begin{document}
\def\ov{\over} \def\be{\begin{equation}} \def\s{\sigma}
\def\ee{\end{equation}} \def\iy{\infty} \def\({\left(} \def\){\right)}  
\def\x{\xi} \def\inv{^{-1}} \def\fh{\hat f} \def\gh{\hat g}
\def\bc{\begin{center}} \def\ec{\end{center}} \def\d{\delta}
\def\R{\mathbb R} \def\sg{{\rm sgn}} \def\G{\Gamma} \def\ph{\varphi} \def\e{\eta} \def\ch{\chi} \def\r{\rho} \def\g{\gamma} \def\fe{\mathfrak{e}}
\def\ep{\varepsilon} \def\ps{\psi} \def\Ph{\Phi} \def\k{\kappa} 
\def\ap{\approx} \def\z{\zeta} \def\iyy{_{-\iy}^\iy} \def\ziy{_0^\iy}
\def\arg{{\rm arg}\,} \def\sgn{{\rm sgn}\,} \def\Re{{\mathcal Re}\,} 
\def\Ft{Fourier transform\ }  \def\ch{\raisebox{.4ex}{$\chi$}} 
\newcommand{\twotwo}[4]{\left(\begin{array}{cc}#1&#2\\&\\#3&#4\end{array}\right)} \def\pmp{{\pm p}} \def\fe{\mathfrak{e}}
\newcommand {\twoone}[2]{\left(\begin{array}{c}#1\\ \\#2\end{array}\right)}
\def\C{\mathcal C} \def\irr{_{-r/2}^{r/2}} \def\wh{\widehat} 
\def\F{\mathcal F} \def\CC{\mathbb C} \def\L1h{\wh{L^1}} \def\I{{\rm Im}\,}
\def\LL{\mathcal L} \def\noi{\noindent} \def\sp{\vspace{1ex}}

\bc{\large\bf On the Ground State Energy\\\vspace{1ex}  of the Delta-Function Fermi Gas}\ec

\bc{\large\bf Craig A.~Tracy}\\
{\it Department of Mathematics \\
University of California\\
Davis, CA 95616, USA}\ec

\bc{\large \bf Harold Widom}\\
{\it Department of Mathematics\\
University of California\\
Santa Cruz, CA 95064, USA}\ec

\begin{abstract} 
The weak coupling asymptotics to order $\gamma$ of the ground state energy of the delta-function Fermi gas, derived heuristically in the literature, is here made rigorous. Further asymptotics are in principle computable. The analysis applies to the Gaudin integral equation a method previously used by one of the authors for the asymptotics of large Toeplitz matrices.

\end{abstract}

\bc{\bf 1. Introduction}\ec

Since Hans Bethe's seminal 1931 paper, the ideas he introduced,   now known under the rubric \textit{Bethe Ansatz}, have had a wide impact in physics and mathematics.  For an historical account
see Batchelor \cite{batchelor} and for book-length treatments  see \cite{gaudinBook, KBI, S2}.  

One of the most widely studied Bethe Ansatz solvable models is the quantum, many-body system in one-dimension with   delta-function two-body interaction \cite{LL}; namely, with
Hamiltonian
\[ H_N= -\sum_{j=1}^N \frac{\partial^2}{\partial x_j^2} + 2 c \sum_{i<j} \delta(x_i-x_j) \]
where $N$ is the number of particles and $2c$ is the coupling constant.
   A basic quantity is  the ground state energy per particle in
the thermodynamic limit:  If $E_0(N,L)$ is the ground state energy for the finite system of $N$ particles on a circle of length $L$,  then in the limit $N\rightarrow \infty$, $L\rightarrow\infty$, such
that $\rho:=N/L$ is fixed, the ground state energy per particle is
\[ \varepsilon_0 := \lim \frac{E_0(N,L)}{N}. \] 
In their now classic paper, Lieb and Liniger \cite{LL}  showed, for particles with Bose statistics and repulsive interaction ($c>0$), that $\fe_B:=\varepsilon_0/\rho^2$ is a function only of $\gamma:= c/\rho$. To
state their basic result, we first define the \textit{Lieb-Liniger operator}
\be \mathcal{L}_\kappa f(x):= \frac{\kappa}{\pi}\int_{-1}^1 \frac{f( y)}{(x-y)^2+\kappa^2}\, dy,\>\>\> -1<x<1. \label{LLop}\ee
Here $\k>0$. If $f_B(x;\k)$ solves the Lieb-Liniger integral equation
\be f(x) -\mathcal{L}_\kappa f(x) =1 ,\label{LLeqn}\ee
then $\fe_B(\gamma)$ is determined, by elimination of $\kappa$,  from the relations
\[ \frac{\kappa}{\gamma}=\frac{1}{2\pi} \int_{-1}^1 f_B(x;\kappa)\,dx, \>\>\> \fe_B(\gamma) = \frac{1}{2\pi} \left(\frac{\gamma}{\kappa}\right)^3 \int_{-1}^1 x^2 f_B(x;\kappa) \, dx. \]
Since $\mathcal{L}_\kappa$ tends to the identity operator as $\kappa\rightarrow 0$, the asymptotic expansion of $\fe_B(\gamma)$ as $\gamma\rightarrow 0$ is nontrivial.  (The asymptotics for $\gamma\rightarrow\infty$ are straightforward.)  It is ``known'' \cite{gaudinBook, LL, popov, TW} that as $\g\to0$,
\[ \fe_B(\gamma) = \gamma - \frac{4}{3\pi} \gamma^{3/2} +\left[\frac{1}{6}-\frac{1}{\pi^2}\right] \gamma^2 +o(\gamma^2).\]
The first two terms in the above expansion have been rigorously justified \cite{gaudinBook} using results of Hutson \cite{hutson}.
 
A natural question   to ask is how the problem changes when the  particles obey Fermi statistics.   The generalization of Bethe Ansatz to this case   was solved  by
 Gaudin \cite{gaudin, gaudinBook} and Yang \cite{yang} (see also \cite{mcg, S1}).  In particular, for spin-1/2 particles with attractive interaction ($c<0$) with total spin zero, the ground state energy per particle in the thermodynamic limit is given by \cite{gaudinBook}
 \[ \frac{\varepsilon_0}{\rho^2} = -\frac{\gamma^2}{4}+\fe_F(\gamma) \]
 where $\gamma = \vert c\vert/\rho$ and  the equation is now the {\it Gaudin integral equation}
  \be f(x) + \mathcal{L}_\kappa f(x) =1.\label{gaudineqn}\ee
 If $f_F(x;\k)$ solves this equation, 
 then $\fe_F(\gamma)$ is determined by elimination of $\kappa$  from the equations
\be \frac{\kappa}{\gamma}=\frac{2}{\pi} \int_{-1}^1 f_F(x;\kappa)\,dx, \>\>\> \fe_F(\gamma) =\frac{2}{\pi} \left(\frac{\gamma}{\kappa}\right)^3 \int_{-1}^1 x^2 f_F(x;\kappa) \, dx. \label{eqns}\ee

The integral equations (\ref{LLeqn}) and (\ref{gaudineqn}) are well-known in the potential theory literature under the name  \textit{Love integral equation} \cite{love,sneddon}.  Equation (\ref{gaudineqn})  also arises in the computation of the charge $Q$
on each
of two coaxial conducting discs of radius one separated by a distance $\kappa$ and each maintained at the same unit potential.  For the Lieb-Liniger equation (\ref{LLeqn}),  the discs are maintained at opposite
potentials $\pm 1$.  In both cases the charge $Q$ is given by the a constant times the zeroth moment of $f$. For the case of equal potentials (the Fermi case), the charge is given by
\be Q={1\ov\pi}\int_{-1}^1 f_F(x;\k)\,dx,\label{Q}\ee
and Leppington and Levine \cite{LepLev} proved rigorously that as $\k\to0$,
\be Q =\frac{1}{\pi} +\frac{\kappa}{2\pi^2} \log\kappa^{-1} +\frac{\kappa}{2\pi^2}\left(\log\pi+1\right)+o(\kappa). \label{Q0}\ee
 The authors did not analyze the integral equation (\ref{gaudineqn}), but rather found an approximate solution of the related boundary value problem.  In later work,
 Atkinson and Leppington \cite{al} analyzed the integral equation  directly by the formal method of matched asymptotic expansions (by finding ``outer'' and ``inner'' solutions).  In this way they reproduced
the result (\ref{Q0}).

Gaudin  used an approximate solution to (\ref{gaudineqn}) (see (11.105) in \cite{gaudinBook}) to conjecture that
\be \fe_F(\gamma) = \frac{\pi^2}{12} - \frac{\gamma}{2} +\textrm{o}(\gamma).\label{eF1}\ee
We say ``conjecture'' since the approximate solution used is not valid near $x=\pm 1$ (as Gaudin himself pointed out); and therefore, the error is not controlled.

Guan and Ma \cite{GM}, using an approximate solution of (4) due to Takahashi \cite{Tak}, deduced the error bound $O(\g^2)$ in (\ref{eF1}). (Although Krivnov and Ovchinnikov \cite{KO} had predicted earlier that the term $\g^2 \log^2\g^{-1}$ appears.) Their method  generalizes to the  ground state energy with a weakly repulsive interaction and with polarization. Using different methods to analyze (\ref{gaudineqn}), Iida and Wadati \cite{iw} found \[ \fe_F(\gamma) = \frac{\pi^2}{12} - \frac{\gamma}{2} +\frac{\gamma^2}{6}+o(\gamma^2).\]
Their analysis was also heuristic since it involved the manipulation of divergent series.  When these same methods were applied to  $\fe_B(\gamma)$, the coefficient of the $\gamma^2$ was in error; but it is uncertain whether this is due simply to an arithmetical error or a more serious error in the method.  (See the discussion in \cite{TW}.)

In this note we provide a rigorous framework for the asymptotics associated with the Fermi gas, based on the integral equation (\ref{gaudineqn}). Although we only verify (\ref{eF1}) by showing that
\be \fe_F(\gamma) = \frac{\pi^2}{12} - \frac{\gamma}{2} +O(\gamma^2 \log^2\gamma^{-1}),\label{eF3}\ee
it is in principle possible (although perhaps tedious) to use the formulas derived here to get further terms in the asymptotics.

Our analysis uses  an analogue of a Wiener-Hopf method used by the second author \cite{w} to derive asymptotics for finite Toeplitz matrices, which are the discrete analogues of finite convolution operators. The main point is that one can replace equation (\ref{gaudineqn}) involving the convolution operator $\LL_\k$ by one involving a pair of Hankel operators for which the Neumann series for the solution is an asymptotic expansion as $\k\to0$.

This method applies to the Fermi gas because, after scaling, the operator $I+\LL_\k$ (here $I$ is the identity operator) has an invertible limit as $\k\to0$. For the operator $I-\LL_\k$ associated with the Bose gas the corresponding limit is not invertible.

In the first two sections we consider a general class of convolution operators and derive the equivalent equations involving a pair of Hankel operators. Then we apply this to the Gaudin operator $I+\LL_\k$. In the following sections we derive the asymptotic results for $Q$ and $\fe_F(\g)$, the former rigorously and the latter not. In the final section we make rigorous the asymptotics for $\fe_F(\g)$.

\bc{\bf 2. Generalities}\ec

Instead of a convolution operator on a fixed interval whose kernel depends on the small parameter $\k$, we consider an operator with fixed kernel on a large interval of length $r$. The two are easliy interchangeable. The convolution equation is 
\[\int_{-r/2}^{r/2}k(x-y)\,f(y)\,dy=g(x),\,\,\,\,-r/2< x< r/2.\] 
Think of $k$, defined on all of $\R$, as having a $\d$-summand and extend $f$ (the unknown function) and $g$ to be zero outside $(-r/2,r/2)$.

We use Fourier transforms. In our notation, the $x\to\x$ Fourier transform has $e^{ix\x}$ in the integrand; the $\x\to x$ inverse \Ft has $e^{-ix\x}$ in the integrand and the factor $1/2\pi$. At first we shall be formal, later more precise. 

If $\s$ is the Fourier transform of $k$, and $\fh$ resp. $\gh$ the \Ft of $f$ resp. $g$, then $\s\fh-\gh$ is the \Ft\ of a function supported outside the interval $(-r/2,r/2)$. The \Ft of a function supported on $(r/2,\iy)$ equals $e^{ir\x/2}$ times the \Ft of a function supported on $\R^+$, the positive reals, and the \Ft of a function supported on 
$(-\iy,-r/2)$ equals $e^{-ir\x/2}$ times the \Ft of a function supported on $\R^-$. Therefore we can write the equation as
\[ \s\fh=\gh+e^{ir\x/2}\,h^++e^{-ir\x/2}\,h^-,\]
where $h^+$ resp. $h^-$ is the \Ft of a function supported on $\R^+$ resp. $\R^-$. We consider these the unknown functions, and we find equations for them. Once they are determined, so is $\fh$.

We denote by $\ps\to \ps_\pm$ the conjugate by the \Ft of multiplication by $\ch_{\R^\pm}$, the charactristic functions of $\R^\pm$. These are given by
\[\ps_\pm(\x)={1\ov2}\ps(\x)\pm{1\ov2\pi i}\,\int\iyy {\ps(\e)\ov \e-\x}\,d\e,\]
where the integral is a principal value. The functions extend analytically to the upper and lower half-planes by the formulas
\[\ps_\pm(\x)=\pm{1\ov2\pi i}\,\int\iyy {\ps(\e)\ov \e-\x}\,d\e,\]
where $\x$ is in the upper half-plane for $\ps_+$ and the lower half-plane for $\ps_-$. 

The {\it Wiener-Hopf factors} of $\s$, which confusingly we denote by $\s_\pm$, are given by
\[\s_\pm=e^{(\log\s)_\pm},\]
where the $\pm$ on the right are the projection operators defined above. For appropriate $\s$ (defined precisely below), $\s_\pm$ and $1/\s_\pm$ equal constants plus Fourier transforms of functions supported on $\R^\pm$. Since the convolution of two functions supported on $\R^\pm$ is also supported on $\R^\pm$, by changing notation we may replace our equation by
\be \s_-\s_+\fh=\gh+e^{ir\x/2}\,\s_+\,h^++e^{-ir\x/2}\,\s_-h^-.\label{eq}\ee

The inverse \Ft of $\fh$ is supported on $(-r/2,\iy)$, so the inverse \Ft of $e^{ir\x/2}\fh$ is supported on $\R^+$, so also the inverse \Ft of $e^{ir\x/2}\s_+\fh$ is supported on $\R^+$. Therefore if we multiply 
(\ref{eq}) by $e^{ir\x/2}/\s_-$ and apply the minus operator we get
\[0=(e^{ir\x/2}\,\s_+\,\fh)_-=\({e^{ir\x/2}\,\gh\ov\s_-}\)_-+\(e^{ir\x}\,{\s_+\ov\s_-}\,h^+\)_-+h^-.\]
Similarly
\[0=\({e^{-ir\x/2}\,\gh\ov\s_+}\)_++h^++\(e^{-ir\x}\,{\s_-\ov\s_+}\,h^-\)_+.\]

Define the operators $U$ and $V$ by 
\[Uu^-=\(e^{-ir\x}\,{\s_-\ov\s_+}\,u^-\)_+,\ \ \ Vv^+=\(e^{ir\x}\,{\s_+\ov\s_-}\,v^+\)_-.\]
The operator $U$ takes Fourier transforms of functions supported on $\R^-$ to Fourier transforms of functions supported on $\R^+$, and $V$ does the opposite. If we define
\be G^-=-\({e^{ir\x/2}\,\gh\ov\s_-}\)_-,\ \ \ G^+=-\({e^{-ir\x/2}\,\gh\ov\s_+}\)_+,\label{G}\ee
our two relations may be written
\[h^-+Vh^+=G^-,\ \ \ h^++Uh^-=G^+.\]
This pair of equations for the functions $h^\pm$ is equivalent to the original equation for the function $f$. The solution is given by
\be\twoone{h^-}{h^+}=\(I+\twotwo{0}{V}{U}{0}\)\inv\,\twoone{G^-}{G^+}
\label{heq}\ee
as long as the inverse on the right side makes sense.

So far everything has been formal. Now we give precise conditions and explain the exact meaning of (\ref{heq}).

Denote by $\L1h$ the Fourier transforms of $L^1$ functions, with the operations and norm inherited from $L^1$, and by $\L1h_\pm$ the Fourier transforms of $L^1$ functions supported on $\R^\pm$.  (In particular convolution on $L^1$ becomes multiplication on $\L1h$.) The operators $\pm$ are projections from $\L1h$ to $\L1h_\pm$. Thus if $\log\s\in\L1h$ then also $(\log\s)_\pm\in\L1h_\pm$, and it follows that the functions $\s_\pm,\ 1/\s_\pm,\ \s_+/\s_-,\,\s_-/\s_+$ all belong to $\CC+\L1h$. If only $\log\s\in\CC+\L1h$ then we modify the definitions of $\s_\pm$ by multiplying by appropriate constants, so the product is $\s$.

Since $\gh\in\L1h$, the functions $G^\pm$ defned by (\ref{G}) belong to $\L1h_\pm$.

We abuse notation temporarily and replace $U$ and $V$ by their conjugates with the Fourier transform, and keep the same notation for the conjugates. If $k_{-/+}$ resp. $k_{+/-}$ is the inverse transform of $\s_-/\s_+$ resp. $\s_+/\s_-$ the operator $U$ has kernel
\[U(x,y)=k_{-/+}(x-y+r),\ \ \ (x\in\R^+,\ y\in\R^-),\]
and $V$ has kernel
\[V(x,y)=k_{+/-}(x-y-r),\ \ \ (x\in\R^-,\ y\in\R^+).\]
After the variable changes $y\to-y$ for $U$ and $x\to-x$ for $V$ these become Hankel operators acting on $L^1(\R^+)$; their kernels are $k_{-/+}(x+y+r)$ and $k_{+/-}(x+y+r)$. Since $\s_+/\s_-$ and $\s_-/\s_+$ belong to $\CC+\L1h$, the functions $k_{-/+}(x)$ and $k_{+/-}(x)$ both belong to $L^1$ away from $x=0$. The norm of $U$ is at most the norm of $k_{-/+}(x+r)$ in $L^1(\R^+)$, which is $o(1)$ as $r\to\iy$. Similarly for $V$. Therefore the inverse on the right side of (\ref{heq}) is well-defined and the Neumann series for the inverse gives an $\L1h$-convergent asymptotic expansion for the left side. 

In particular we get convergent asymptotic expansions for $h^\pm(0)$, since the linear functional $h\to h(0)$ is continuous on $\L1h$. For the Fermi gas, this will be enough to get the terms in an aymptotic expansion for $\fh(0)$ and so for $Q$. We can also find the terms in an asymptotic expansion for $\fe_F(\g)$, but this will not be rigorous at first. The reason is that the coefficients of the higher powers of $\x$ in the asymptotic expansion of $h^\pm$ (and therefore the higher moments of $f$) are not continuous in the norm of $\L1h$. To justify these asymptotics we will need more properties of the operators $U$ and $V$, among other things. This will be done in Section 7.

\bc{\bf 3. The Fermi gas}\ec

To avoid factors of $\sqrt2$ later we consider the integral equation 
\[{f(x)\ov2}+{1\ov2\pi}\int_{-r/2}^{r/2}{f(y)\ov (x-y)^2+1}\,dy=1,\]
so that
\[\s(\x)=(1+e^{-|\x|})/2,\ \ \ \gh(\x)=2\,\sin(r\x/2)/\x.\]

The $\k$ in (\ref{LLop}) and $r$ are related by $r=2/\k$. The solution $f_F(x;\k)$ of (\ref{gaudineqn}) and our $f(x)$ are related by $f(rx/2)=2\,f_F(x;\k)$. From (\ref{Q}) we get
\be Q={1\ov r\pi}\,\int_{-r/2}^{r/2}f(x)\,dx={\k\ov2\pi}\,\int_{-r/2}^{r/2}f(x)\,dx.\label{Q1}\ee
From the first part of (\ref{eqns}) we find that
\be\g=\k\,\({2\ov\pi r}\int_{-r/2}^{r/2} f(x)\,dx\)\inv={1\ov2}\k \,Q\inv.\label{gamma}\ee
Since $\s_\pm(0)=1$ and $\gh(0)=r$, we have 
\be\int_{-r/2}^{r/2}f(x)\,dx=\fh (0)=r+h^+(0)+h^-(0),\label{rhh}\ee
which by (\ref{Q1}) determines $Q$ from $h^\pm(0)$. From both parts of (\ref{eqns}) and a little computation we find that
\be\fe_F(\g)=\pi^2{{\displaystyle\int_{-r/2}^{r/2} x^2\,f(x)\,dx\ov\(\displaystyle\int_{-r/2}^{r/2} f(x)\,dx\)^3}}.\label{e}\ee

The functions in (\ref{G}) are here
\be G^-=i\,\({e^{ir\x}-1\ov\x\s_-}\)_-,\ \ \ G^+=i\,\({1-e^{-ir\x}\ov\x\s_+}\)_+,\label{G1}\ee
and the Wiener-Hopf factors are given by
\[\log\,\s_\pm(\x)=\pm{1\ov2\pi i}\int\iyy {\log\((1+e^{-|\e|})/2\)\ov \e-\x}\,d\e.\] 
For $\x\in\R$ these become the limits as $\x\to\R$ from above and below $\R$.

It is easy to see that the inverse \Ft of $\log(1+e^{-|\x|})$ belongs to $L^1$. It follows, as discussed earlier, that the functions $\s_\pm,\ 1/\s_\pm,\ \s_+/\s_-,\,\s_-/\s_+$, restricted to $\R$, all belong to $\CC+\L1h$.

It is also easy to see that $\s_\pm(\x)$ and $1/\s_\pm(\x)$ extend boundedly and analytically to any substrip of $|\I \x|<\pi$ cut along the imaginary axis. Explicitly, we have \cite{al}
\[\s_+(\x)=\pi^{1/2}\,\exp\left\{{\x\ov2\pi i}\Big[\log(-i\x)-\log2\pi-1\Big]\right\}\,\G\({1\ov2}+{\xi\ov2\pi i}\)\inv,\]
\[\s_-(\x)=\pi^{1/2}\,\exp\left\{-{\x\ov2\pi i}\Big[\log(i\x)-\log2\pi-1\Big]\right\}\,\G\({1\ov2}-{\xi\ov2\pi i}\)\inv.\]
For $\x$ in the upper resp. lower half-plane, $-i\x$ resp. $i\x$ lies in the right half-plane and the principal values of the logarithms are taken.

We  first find bounds for the norms of the operators $U$ and $V$ better than the bound $o(1)$ which holds generally. As discussed at the end of the last section, the kernels of the conjugates of the operators by the \Ft are given in terms of the inverse Fourier transforms of $\s_-/\s_+$ and $\s_+/\s_-$ for $x>0$ and $x<0$, respectively. The inverse \Ft of $\s_-/\s_+$ is equal to the $\d$-function plus $1/2\pi$ times
\be\int\iyy e^{-ix\x}\({\s_-(\x)\ov\s_+(\x)}-1\)\,d\x,\label{Ft1}\ee
and here $x>0$. The function $\s_-(\x)$ is analytic in the lower half-plane. 

We use $\s_-(\x)/\s_+(\x)=\s_-(\x)^2/\s(\x)$. The function $1/\s(\x)$ extends extends boundedly and anlytically to the lower half-plane cut along the negative imaginary axis. We swing the $\R^+$ part of the contour clockwise down to the right side of the imaginary axis and the $\R^-$ part of the contour counterclockwise down to the left side of the imaginary axis. There we make the substitution $\x=-iy$. The analytic continuation of $1/\s(\x)$ to the right side of the imaginary axis minus its analytic continuation to the left side of the imaginary axis equals 
\[{2\ov 1+e^{iy}}-{2\ov 1+e^{-iy}}=-2i\tan(y/2).\]
Therefore the integral equals.  
\be-{1\ov\pi}\int_0^\iy e^{-xy}\s_-(-iy)^2\,\tan(y/2)\,dy.\label{Ft2}\ee
(This is a prinipal value integral at each odd multiple of $\pi$. The contributions of the integrals over the little semicircles on either side of the imaginatry axis cancel each other.)
The non-exponential part of the integarnd is $O(y)$ as $y\to0$, so the  integral itself is $O(x^{-2})$ as $x\to\iy$.

The kernel of $U$, an operator from $L^1(\R^-)$ to $L^1(\R^+)$, is $k_{-/+}(x-y+r)$. We have shown that $k_{-/+}(x)=O(x^{-2})$ for $x>0$. Therefore
\[\|U\|\le\int_0^\iy |k_{-/+}(x+r)|\,dx=O(r\inv).\]
Similarly $\|V\|=O(r\inv)$.

\bc{\bf 4. Expansion of \boldmath$G^\pm(\x)$ near \boldmath$\x=0$}\ec

The expression for $G^+(\x)$ for $\x$ in the upper half-plane is
\[G^+(\x)={1\ov2\pi}\int\iyy{1-e^{-ir\e}\ov\e\,\s_+(\e)}\,{d\e\ov\e-\x}.\]

We push the contour up to $\R+ic$ with $c>{\rm Im}\, \x$, pass the pole at $\e=\x$, and obtain
\[i{1-e^{-ir\x}\ov\x\,\s_+(\x)}-{1\ov2\pi}\int_{ic-\iy}^{ic+\iy}{e^{-ir\e}\ov\e\,\s_+(\e)}\,{d\e\ov\e-\x}.\]
(The reason the the summand 1 does not appear in the integrand is that the contour for the integral with it can be pushed all the way up.) The integral can be expanded in powers of $\x$ near $\x=0$, and the above becomes
\be i\,{1-e^{-ir\x}\ov\x\,\s_+(\x)}-{1\ov2\pi}\sum_{k=0}^\iy \x^k \int_{ic-\iy}^{ic+\iy}{e^{-ir\e}\ov \s_+(\e)}\,\e^{-k-2}\,d\e=
i\,{1-e^{-ir\x}\ov\x\,\s_+(\x)}+\sum_{k=0}^\iy g_k^+\,\x^k.\label{Grep}\ee

Near zero there is an expansion, valid for $|\x|<\pi$,
\[{1\ov\s_+(\x)}=\sum_{0\le m\le n}a_{n,m}\,\x^n\,\log^m(-i\x),\]
with $a_{0,0}=1$. So the first term in (\ref{Grep}) is equal to
\[i\,(1-e^{-ir\x})\,\sum_{0\le m\le n}a_{n,m}\,\x^{n-1}\,\log^m(-i\x).\]

To determine the asymptotics of the coefficients 
\[g_k^+=-{1\ov2\pi}\int_{ic-\iy}^{ic+\iy}{e^{-ir\e}\ov \s_+(\e)}\,\e^{-k-2}\,d\e\]
 as $r\to\iy$ we do what we did for the \Ft of $\s_-/\s_+$, with some change. Now we make sure that $c\in(0,\pi)$ and deform the contour to the one from $-ic-\iy$ to $-ic-0$, up the left side of the imaginary axis, around a circle with center zero (say with radius $c/2$), then down the right side of the imaginary axis to $-ic+0$, and finally to $-ic+\iy$.
The integral over the the horizontal parts of the contour are exponentially small as $r\to\iy$. If we take only the terms with $n\le N$ in the series for $1/\s_+(\e)$ the error is $O(\e^N)$ near $\e=0$. Then we can shrink the circle around zero, getting an integral over portions of the imaginary axis of the form (after making the substitution $\e=-iy$)
\[\int_0^c e^{-ry}\,O(y^{N-k-2})\,dy.\]
This is $O(r^{-N+k+1})$. 

It follows that for an asymptotic expansion of $g_k^+$ as $r\to\iy$ we may replace $1/\s_+(\x)$ by the sum for it and then interchange the sum over $n,m$ with the integral. The contour goes around the circle and over portions of the imaginary axis. For each summand we may complete the contour by adding the vertical lines from  $-ic\pm0$ to $-i\iy\pm0$, incurring only an exponentiall small error. Think of the resulting contour $\C$ as starting from $-i\iy-0$, looping clockwise around zero, and then down to $-i\iy+0$. 

We have shown that an asymptotic expansion for $g_k^+$ is given by
\[g_k^+\sim -{1\ov2\pi}\sum_{0\le m\le n}a_{n,m}\int_\C e^{-ir\e}\,(-i\e)^{n-k-2}\,\log^m(-i\e)\,d\e.\]
To identify the integral we consider first the integral
\[\int_\C e^{-ir\e}\,(-i\e)^{-s-1}\,d\e=r^{s}\,\int_\C e^{-i\e}\,(-i\e)^{-s-1}\,d\e,\]
with $s$ a continuous variable and the specification $|\arg(-i\e)|<\pi$.  
If we make the substitution $\e\to -i\e$ then this becomes
\[-i\,r^{s}\,\int_{i\C} e^{-\e}\,(-\e)^{-s-1}\,d\e.\]
The contour $i\C$ loops around the positive real axis clockwise. The integral is recognized as one for the reciprocal of the gamma function \cite[\S12.22]{ww}, and we see that the above equals $2\pi\,r^{s}\,\G(s+1)\inv$.

If we differentiate $m$ times with respect to $s$ we obtain
\[\int_\C e^{-ir\e}\,(-i\e)^{-s-1}\,\log^m(-i\e)\,d\e=2\pi\,\(-{d\ov ds}\)^m{r^{s}\ov\G(s+1)},\]
and so
\[\int_\C e^{-ir\e}\,\e^{n-k-2}\,\log^m(-i\e)\,d\e=-2\pi\,(-i)^{k-n}\,\(-{d\ov ds}\)^m{r^{s}\ov\G(s+1)}\Big|_{s=k-n+1}.\]

Putting things together gives the small-$\xi$ expansion
\[G^+(\x)=i\,(1-e^{-ir\x})\,\sum_{0\le m\le n}a_{n,m}\,\x^{n-1}\,\log^m(-i\x)\]
\be+\sum_{k=0}^\iy (-i\x)^k\,\sum_{0\le m\le n}\,i^{n}\,a_{n,m}\,\(-{d\ov ds}\)^m{r^{s}\ov\G(s+1)}\Big|_{s=k-n+1},\label{Gexpand}\ee
where the sum multiplying $(-i\x)^k$ is an asymptotic expansion as $r\to\iy$ for its coefficient. 

\noi{\bf Remark.} An easy check shows that the sum of terms with fixed $n$ and $m=0$ equals zero when $n=0$ and $i\,\x^{n-1}\,a_{n,0}$ when $n>0$.
In particular, only terms with $n>0$ contribute to the sum.

The coefficient of $\x^k$ is of order $r^k$. Any term with $n>1$ will contribute to this coefficient at most $r^{k-1}\log^2 r$. So for a first approximation we may take $n=1$ only. When $m=0$ we use the Remark.
We have (here $\g$ is the Euler $\g=-\G'(1)$) 
\[a_{1,0}={i\ov2\pi}(\g-\log(\pi/2)-1),\ \ a_{1,1}={i\ov2\pi},\]
and using this and (\ref{Gexpand}) we find the following coefficients of $\x^k$ up to $k=2$, with error $r^{k-1}\log^2 r$:

\[\quad\quad k=0:\ \ i\,a_{1,0}-i(\log r+\g)\,a_{1,1}={1\ov2\pi}(\log r+\log(\pi/2)+1);\]
\[\! k=1:\ \ -a_{1,1}\,r\,(\log r+\g-1)=-{i\ov2\pi}r\,(\log r+\g-1);\] 
\[\quad\quad k=2:\ \ i\,a_{1,1}\,{r^2\ov2}\,(\log r+\g-3/2)=-{1\ov4\pi}r^2\,(\log r+\g-3/2).\]

In particular,
\be G^+(0)={1\ov2\pi}(\log r+\log(\pi/2)+1)+O(r\inv\log^2 r).\label{G0}\ee

{\bf Remark.} The coefficients above come from the last sum in (\ref{Gexpand}). The $n=m=1$ term from the other sum does not contribute to the constant term, and its eventual contribution to the $\x^2$ term on the right side of (\ref{eq}) will be of lower order and so may be ignored. 
\pagebreak

\bc{\bf 5. Calculation of \boldmath$Q$}\ec

It follows from the fact that the norms of $U$ and $V$ are $O(r\inv)$ that
\be|h^+(0)-G^+(0)|\le\|h^+-G^+\|=O(r\inv\,\|G^+\|),\label{hminusG}\ee
the last norm being that in $\L1h$. This is the same as the $L^1$-norm of the inverse \Ft of $G^+$. This inverse \Ft is the convolution of the inverse \Ft of
$i(1-e^{-ir\x})/\x$, which is minus $\ch_{(-r,0)}$, and the inverse \Ft of 
$1/\s_+(\x)$. If we denote the latter by $s(x)$ then the convolution at~$x$ equals 
\[\int_x^{x+r}s(y)\,dy.\]
The norm of this in $L^1(\R^+)$ is bounded by
\[\int_0^\iy \min(r,y)\,|s(y)|\,dy.\]
Since $1/\s_+\in\CC+\L1h$, we know that $s(y)$ is a linear combination of a $\d$-function, which drops out, and an $L^1$ function. Therefore the integral over $(0,1)$ is bounded. For the rest, an argument analogous to the one in which we bounded the \Ft of $\s_-/\s_+$ shows that $s(y)=O(y^{-2})$. One sees from this that the integral over $(1,\iy)$ is $O(\log r)$. Hence (\ref{hminusG}) is $O(r\inv\log r)$, and with this and (\ref{G0}) we have shown
\[h^+(0)={1\ov2\pi}(\log r+\log(\pi/2)+1)+O(r\inv\log^2 r).\]

Similarly for $h^-(0)$. Therefore from (\ref{rhh}) and (\ref{Q1}) we obtain
\be \int\irr f(x)\,dx=\fh(0)=r+{1\ov\pi}(\log r+\log(\pi/2)+1)+O(r\inv\log^2 r),\label{fint}\ee
\[Q={1\ov\pi}+{\k\ov2\pi^2}\log\k\inv+{\k\ov2\pi^2}(\log\pi+1)+O(\k^2\log^2\k\inv).\]
This agrees with (\ref{Q0}). 

\bc{\bf 6. Calculation of \boldmath$\fe_F(\g)$}\ec

It follows from (\ref{fint}) that
\be\(\int\irr f(x)\,dx\)^{-3}=r^{-3}\,\Big(1-3\pi\inv(\log r+\log(\pi/2)+1)\,r\inv+O(r^{-2}\log^2 r)\Big).\label{intcubed}\ee
This gives the denominator in (\ref{e}). For the numerator is easy to see that
\[\int\irr x^2\,f(x)\,dx=-\fh''(0)\]
\be=\fh(0)/2-2\times \textrm{the coefficient of $\x^2$ in the expansion of}\ \s(\x)\fh(\x).\label{moment}\ee
The first term is $O(r)$, as we know, and will be of lower order than the rest. The $\gh(\x)$-term in (\ref{eq}) gives to the coefficient the contribution $-r^3/24+O(r)$, so we have to consider the $h^\pm$ terms. {\it Now we assume that the coefficients in the expansion of $G^\pm$ are first-order approximations to the corresponding coefficients for $h^\pm$.}

From (\ref{eq}) we must multiply $G^+(\x)$ by $e^{ir\x/2}$ and take the coefficient of $\x^2$. (We may ignore the factor $\s_+(\x)$ because it would only contribute extra powers of $\x$ without accompanying powers of $r$.) From the coefficients  of $G^+(\x)$ we computed we find that the coefficient of $\x^2$ in $e^{ir\x/2}\,G^+(\x)$ is
\[-{r^2\ov16\pi}(\log r+\log(\pi/2)-1)+O(r\log^2 r).\]

We double this (there is an equal term from the $h^-$ summand), add the term $-r^3/24$ from the $\gh$-term in (\ref{eq}), and multiply by (\ref{intcubed}). The result is
\[-{1\ov24}\(1-{6\ov \pi r}\)+O(r^{-2}\log^2r)=-{1\ov24}\(1-{6\g\ov\pi^2}\)+O(\g^{2}\log^2\g\inv),\]
since $\g=\pi/r+O(\log^2 r/r^2)$. (See (\ref{gamma}) and the asymptotics of $Q$.) Then multiplying by $-2$ from (\ref{moment}) and $\pi^2$ from (\ref{e}) give
\[\fe_F(\g)={\pi^2\ov12}-{\g\ov2}+O(\g^{2}\log^2\g\inv).\]

To make this rigorous we have to show that the coefficients in the asymptotic expansions of $h^\pm$ are, to a first order, equal to the corresponding coefficients in the asymptotic expansions of $G^\pm$. This, and more, will be done in the next section.

\bc{\bf 7. Making the asymptotics rigorous}\ec

In the following lemmas, when we consider the result of the operators $\ps\to\ps_\pm$ the function $\ps_\pm$ will mean its analytic extension to the corresponding half-plane. 

We use the following notation: A bound $O(r^{n+})$ as $r\to\iy$ means $O(r^n\log^mr)$ for some $m\ge0$. Analogously, $O(\x^{n+})$ as $\x\to 0$ means 
$O(|\x|^{n}\log^m|\x|\inv)$ for some $m\ge0$. This notation makes it unnecessary to keep track of logarithm factors in bounds.

We define $\LL^+$ to be those families of functions $\ph\in\L1h_+$ (Fourier transforms of $L^1$ functions supported on $\R^+$), depending on the parameter $r$, satisfying the following:

\noi(i) $\|\ph\|=O(r^{0+})$ as $r\to\iy$;

\noi(ii) There is an asymptotic expansion as $\x\to0$,
\[\ph(\x)\sim \sum_{0\le m\le n}c_{n,m,r}\,\x^n\,\log^m(-i\x),\]
where each $c_{n,m,r}=O(r^{n+})$ as $r\to\iy$; and stopping the series with terms $n<N$ leads to an error $O((r\x)^{N+})$.
 
Similarly we define $\LL^-$. It is easy that $\LL^+$ is closed under multiplication by $e^{\pm ir\x/2}$ or $\s_+(\x)^{\pm1}$. This will be relevant when we go back to (\ref{eq}). In the following lemma $r\LL^+$ denotes $r$ times the functions in $\LL^+$. Analogous notation will be used later.
\sp

\noi{\bf Lemma 1}. We have $\gh_+,\,(\s\,\fh)_+\in r\LL^+$.
\sp

\noi{\bf Proof}. The statement concerning $\gh_+$ is almost immediate. As for $\fh$, if $K$ is the operator with convolution kernel then $f=(I+K)\inv1=(I-K^2)\inv\,(I-K)1>0$ since $(K1)(x)<1$. Since $f+Kf=1$, it follows that $0< f<1$, so $\|\fh\|=\|f\|<r$. Therefore $\|\s\fh\|=O(r)$, and then $\|(\s\fh)_+\|=O(r)$. Thus (i), with the extra factor $r$, is satisfied.

For (ii) we write $\fh=\fh_++\fh_-$ and first consider $(\s\,\fh_-)_+$. For $\x$ in the upper half-plane 
\[(\s\,f_-)_+(\x)={1\ov2\pi i}\int\iyy \s(\e)\,f_-(\e)\,{d\e\ov\e-\x}.\]
We do something similar to what we did before, namely swing the $\R^+$ part of the contour down to the right part of the imaginary and the $\R^-$ part of the contour down to the left part of the imaginary axis. Then we set $\e=-iy$. The continuation of $\s(\e)$ to the right part of the imaginary axis minus 
continuation of $\s(\e)$ to the left part of the imaginary axis equals $i\sin y$, so the integral becomes
\[{1\ov2\pi}\int_0^\iy \fh_-(-iy)\,\sin y \,{dy\ov y-i\x},\]
and $-i\x$ is in the right half-plane. We rewrite this as
\be{1\ov2\pi}\int_0^{r/2}f(-x)\,dx\,\int_0^\iy e^{-xy}\,\sin y \,{dy\ov y-i\x}.\label{fint1}\ee

For the inner integral we use the general formula (\ref{result1}) below in the limit $s\to0$, with $\x$ replaced by $-\x$ and $r$ by $x':=x\pm i$, to obtain
\[\int_0^\iy e^{-x'y}\,{dy\ov y-i\x}=-e^{-ix'\x}\,\log(-ix'\x)+\sum_{n=0}^\iy{(-ix'\x)^n\ov n!}\,\ps(n+1),\]
where $\ps=\G'/\G$. Then we replace $x'$ by $x\pm i$, multiply by $f(-x)$, integrate, subtract the two results, and divide by $2i$. We find a series of the form in (ii) with bounds on the coefficients that are required.

For what comes below, we note that the $\log(-i\x)$ terms in the expansion of the inner integral in (\ref{fint1}) combine as
\be-{e^{-ix\x}\ov 2i}\,(e^{-\x}-e^\x)\,\log(-i\x).\label{logterm1}\ee

The error after stopping the power series above at $\x^{N-1}$ 
is $O((x\x)^N)$, and stoppoing the series for the logarithn term there is $O((x\x)^N\log x)$. After mutiplying by $f(-x)$ and integrating over $(0,r/2)$ we get the bound $O(\x^N r^{N+1}\log r)$ for the error. This will give the rest of (ii), with the extra factor $r$.

We are not finished beacause we want $(\s\fh)_+$ and what we considered was $(\s\fh_-)_+$. That leaves 
\[(\s\fh_+)_+(\x)={1\ov2\pi i}\int_0^{r/2}f(x)\,dx\int\iyy e^{ix\e}\,\s(\e){d\e\ov\e-\x}.\]
For the inner integral we deform to the imaginary axis, and when we swing the $\R^+$ part of the contour we use $\s(\e)=(e^{-\e}+1)/2$ and when we swing the $\R^-$ part of the contour we use $\s(\e)=(e^{\e}+1)/2$. Suppose for definiteness that $\x$, which is in the upper half-plane, is also in the right half-plane. Then we pass the pole at $\e=\x$ when we swing the $\R^+$ part of the contour and find that the inner integral becomes (after the substitution $\e=iy$ on the positive imaginary axis) 
\[\pi i \,e^{ix\x}\,(e^{-\x}+1)-i\int_0^\iy e^{-xy}\,\sin y\,{dy\ov y+i\x}.\]

The integral here is the inner integral in (\ref{fint1}) with $\x$ replaced by $-\x$. Thus the integral with its factor equals a power series in $\x$ (with the bound after stopping at powers of $\x$ less than $N$) plus the logarithm term. This is obtained from (\ref{logterm1}) after the substitution $\x\to-\x$ and multiplying by the factor $-i$, with the result
\be{e^{ix\x}\ov 2}\,(e^{\x}-e^{-\x})\,\log(i\x).\label{logterm2}\ee

How does $\log(i\x)$ relate to $\log(-i\x)$, the logarithm  analytic for all $\x$ in the upper half-plane? All our logarithms are principal value. Thus for $\x$ in the upper half-plane $\log(-i\x)$ is determained by $|\arg(-i\x)|<\pi/2$. With $\x$ in the first quadrant, as we have been taking it, we have $\pi/2<\arg(i\x)<\pi$ whereas $-\pi/2<\arg(-i\x)<0$. Thus $\log(i\x)=\log(-i\x)+i\pi$. It follows that the term before the integral plus (\ref{logterm2}) equals
\[i\pi\,e^{ix\x}(e^{-\x}+1)+i{\pi\ov2}\,e^{ix\x}\,(e^{\x}-e^{-\x})+{e^{ix\x}\ov 2}\,(e^{\x}-e^{-\x})\,\log(-i\x)\]
\[={i\pi\ov 2}\,e^{ix\x}\,(e^\x+e^{-\x}+2)+{e^{ix\x}\ov 2}\,(e^{\x}-e^{-\x})\,\log(-i\x).\]
Although we assumed that $\x$ is in the right half-plane, we now have a function that extends analytically to the entire upper half-plane, because $\log(-i\x)$ does, so the same identity holds throughout this half-plane. We then multiply by $f(x)$ and integrate over $(0,r/2)$. The result satisfies condition~(ii), with the extra factor $r$ coming from the first summand. We then continue as before with the power series.\hfill$\Box$
\sp

\noi{\bf Lemma 2}. If $\ph\in\LL^+$ then $V\ph\in r\inv\,\LL^-$, and if $\ph\in\LL^-$ then $U\ph\in r\inv\,\LL^+$.
\sp

\noi{\bf Proof}. We derive the first statement. For (i) we use that $\|V\|=O(r\inv)$. For (ii), an argument analogous to the one that leads from (\ref{Ft1}) to (\ref{Ft2}) gives
\[V\ph(\x)=-{1\ov\pi}\int_0^\iy e^{-ry}\,\s_+(iy)^2\,\ph(iy)\,\tan(y/2){dy\ov y+i\x}.\]
Here $\x$ is in the lower half-plane, so $i\x$ is in the right half-plane. 
We have $\ph(iy)=O(r)$ uniformly in $y$ by (i). The integral over $(1,\iy)$ equals a power series in $\x$ each of whose coefficients is exponentially small in $r$. The error after stopping the series at $\x^{N-1}$ is bounded by $\x^N$ times an exponentially small factor. So we may replace the interval of integration by $(0,1)$.

If we truncate the series for $\s_+(iy)^2$ at all powers of $y$ less than $N$ the error is $O(y^{N+})$. For $\ph(iy)$ the error is $O((ry)^{N+})$, by assumption (ii).  Therefore, because of the factor $\tan(y/2)$, if we truncate the series for $\s_+(iy)^2\,\ph(iy)$ at all powers of $y$ less than $N$ the error is $O(r^N y^{N+1+})$. With the error only, we use
\[{1\ov y+i\x}=\sum_{n=0}^N (-i\x)^n\,y^{-n-1}+O(\x^N\,y^{-N-1}).\]
For $n<N$ the $n$th term of the series, after multiplying by the error and integrating, gives $\x^n$ times $O(r^{n-1+})$. So the integral with the error term instead gives an expansion with powers of $\x$ up to $N-1$ satisfying condition (ii) with an extra factor $r\inv$. It follows that to extablish (ii), with this extra factor, for the full integrand it suffices to establish it for each summand of the integrand. And for this we may go back to the interval of integration $(0,\iy)$.

So consider the integral when $\s_+(iy)^2\,\ph(iy)\,\tan(y/2)$ is replaced by $y^{n+1}\,\log^m y$ with $m\le n$, wich may come from one term of its expansion. (The extra power of $y$ comes from the factor $\tan(y/2)$.) The coefficient is $O(r^{n+})$. We use
\[\int_0^\iy e^{-ry}\,y^s\,{dy\ov y+i\x}={\pi\ov\sin\pi s}\int_{i\C}e^{-ry}\,(-y)^s\,{dy\ov y+i\x},\]
where as in section 4 the contour $i\C$ loops around the positive real axis clockwise. The pole at $-i\x$ is outside the contour. If we expand the contour to pass the pole we get a contribution from the residue at the pole, and the integrand in the resulting integral can be expanded in powers of $\x$. The coefficients are, as in section~4, integrals for the reciprocal of the gamma function. The result is
\be\int_0^\iy e^{-ry}\,y^s\,{dy\ov y+i\x}={\pi\ov\sin\pi s}\left[-e^{ir\x}\,(i\x)^s+r^{-s}\sum_{k=0}^\iy{(ir\x)^k\ov\G(k-s+1)}\right].\label{result1}\ee
For $s=n+1$ the expression in brackets vanishes and the integral equals $(-1)^{n+1}$ times the derivative with respect to $s$ of the expression in brackets, evaluated at $s=n+1$. The error after stopping at powers of $\x$ less than $N$ is $O(r^{-n-1+}(r\x)^N)=O(r^{-1+}(r\x)^N)$, which is part of (ii) with the extra factor $r\inv$. To get
\[\int_0^\iy e^{-ry}\,y^s\,\log^m y{dy\ov y+i\x}\]
with $m\le n$ we first differentiate $m$ times with respect to $s$. 

Consider the coefficient of $\x^k$, with $k\ne n+1$, coming from the series. After differentiating $m$ times with respect to $s$ and setting $s=n+1$ the coefficient is at most $O(r^{k-n-1}\log^m r)$. Multiplying the by the original coefficient $O(r^{n+})$ gives $O(r^{k-1+})$. 

For the term involving $\x^{n+1}$ we get a factor of at most $\log^{m+1} (i\x)$, so at most $\log^{n+1} (i\x)$. For the coefficient, we get at most $O(\log^{m+1}r)$ which, when multiplied by the original coefficient $O(r^{n+})$ gives $O(r^{n+})=(r\inv\,r^{n+1+})$. 

This verifies the rest of (ii) with the extra factor $r\inv$. \hfill$\Box$
\sp

\noi{\bf Lemma 3}. We have $G^\pm\in\LL^\pm$.
\sp

\noi{\bf Proof}. We already know that $\|G^+\|=O(\log r)$. For (ii), we write
\[2\pi \,G^+(\x)=\int\iyy{1\ov\e\,\s_+(\e)}\,{d\e\ov\e-\x}-\int\iyy{e^{-ire}\ov\e\,\s_+(\e)}\,{d\e\ov\e-\x}\]
\[={2\pi i\ov\x\,\s_+(\x)}-\int\iyy{e^{-ir\e}\ov\e\,\s_+(\e)}\,{d\e\ov\e-\x}.\]
We deform the contour in the integral to one going up the left side of the imaginary axis, then in a little circle around zero, then down the right side of the imaginary axis. The integral around the little circle becomes in the limit $2\pi i/\x$,
and combined with the first term gives
\[{2\pi i\ov\x}\({1\ov\s_+(\x)}-1\)=\log(-i\x)+{\rm const.} +o(1),\] 
near zero.

After using $1/\s_+=\s_-/\s$ the integral over the negative imaginary axis becomes, as in the proof of Lemma 2,
\[\int_0^\iy {e^{-ry}\ov y}\,\s_-(iy)\,\tan(y/2){dy\ov y-i\x},\]
and this is handled as in the proof of that lemma. The only difference is the denominator $y$ which gives a $\log(-i\x)$ summand, which cancels the $\log(-i\x)$ we had above, and because of which the exrtra factor $r\inv$ does not occur. We omit the details.\hfill$\Box$

\noi{\bf Lemma 4}. For each $k\ge0$,
\be\twoone{h^-}{h^+}-\sum_{j=0}^{k-1}(-1)^j\,\twotwo{0}{V}{U}{0}^j\,\,\twoone{G^-}{G^+}\in r^{-k}(\LL^-\oplus\LL^+).\label{result}\ee

\noi{\bf Proof}. It follows from Lemma 1, by applying the plus operator to (\ref{eq}), that the function $e^{ir\x/2}\,\s_+(\x)\,h^+(\x)$ belongs to $r\LL^+$. Therefore the same is true of $h^+(\x)$, and similarly $h^-(\x)$. From this, the representation (\ref{heq}), and Lemma 2 we deduce that for each $k$
\[\twotwo{0}{V}{U}{0}^{k+1}\(I+\twotwo{0}{V}{U}{0}\)\inv\,\twoone{G^-}{G^+}\in r^{-k}(\LL^-\oplus\LL^+).\]
By the general fact $L^{k+1}\,(I-L)\inv=(I-L)\inv-\sum_{j=0}^k L^j$, this is the same as
\[\twoone{h^-}{h^+}-\sum_{j=0}^{k}(-1)^j\,\twotwo{0}{V}{U}{0}^j\,\,\twoone{G^-}{G^+}\in r^{-k}(\LL^-\oplus\LL^+).\]
But by Lemmas 2 and 3 we know that
\[\twotwo{0}{V}{U}{0}^k\,\,\twoone{G^-}{G^+}\in r^{-k}(\LL^-\oplus\LL^+).\]
Multiplying this by $(-1)^k$ and adding gives the statement of the lemma.\hfill$\Box$
\sp

Formula (\ref{Gexpand}) gives the the asymptotic expansion of $G^+$, and therefore also for $G^-$ since functions are complex conjugates of each other. It follows from Lemma~4 that the coefficient of $\x^n$ in $h^\pm$ is obtained with error $O(r^{n-k+})$ by taking the corresponding coefficient in the sum in (\ref{result}) up to $k-1$. With the formulas in the proof of Lemmas~2 and~3 further coefficients can be computed from the asymptotics of some integrals. And then we would determine the coefficients in the expansions of their products with $e^{\pm ir\x/2}\,\s_\pm(\x)$ appearing in (\ref{eq}). Thus with the result for general $k$  one could in principle find as may terms as desired in the asymptotic expansions  of $Q$ and $\fe_F(\g)$.

The special case $n=2,\ k=1$ justifies the asymptotics we found for $\fe_F(\g)$. Because we did not keep track of powers of logarithms, we cannot guarantee that the power two of the logarithm in the error term in (\ref{eF3}) is correct, only that it is correct with some power.

\bc{\bf Acknowledgments}\ec

We thank the referee for pointing out refreences \cite{GM} and \cite{KO}.

This work was supported by the National Science Foundation through grants DMS--1207995 (first author) and DMS--1400248 (second author).


\begin{thebibliography}{99}

\bibitem{al} C.~Atkinson and F.~G.~Leppington,  The asymptotic solution of some integral equations, IMA J.\ Appl.\  Math.\  {\bf 31} (1983) 169--182.

\bibitem{batchelor} M.~T.~Batchelor, The Bethe ansatz after 75 years, Physics Today, \textbf{60} (2007) , 36--40.

\bibitem{gaudin} M.~Gaudin, Un systeme \`a une dimension de fermions en interaction, Phys.\ Lett. A, \textbf{24} (1967), 55--56.

\bibitem{gaudinBook} M.~Gaudin, \textit{The Bethe Wavefunction}, Cambridge 
Univ.\ Press, 2014. (English translation by J.-S.~Caux.)

\bibitem{GM} X-W. Guan and Z-Q. Ma, One-dimensional multicomponent fermions with $\delta$-function interaction in strong- and weak-coupling limits: Two-component Fermi gas, Phys.\ Rev.\ A \textbf{85} (2012), 033632.

\bibitem{hutson} V.~Hutson, The circular plate condenser at small separation, Proc.\ Camb.\ Phil.\ Soc.\ \textbf{59} (1963), 211--224.

\bibitem{iw} T.~Iida and M.~Wadati, Exact analysis of a $\delta$-function spin-1/2 attractive Fermi gas with arbitrary polarization,
J.\ Stat.\ Mech.\ (2007) P06011.

\bibitem{kw} T.~Kaminaka and M.~Wadati, Higher order solutions of the Lieb-Liniger integral equation,
Phys.\ Lett. A \textbf{375} (2011), 2460--2464.

\bibitem{KBI} V.~E.~Korepin, N.~M.~Bogoliubov, and A.~G.~Izergin, \textit{Quantum Inverse Scattering Method and Correlation Functions},
Cambridge Univ.\ Press, 1993.

\bibitem{KO} V. Ya. Krivnov and A. A. Ovchinnikov, One-dimensional Fermi gas with attarction between the electrons, Sov. Phys. JETP {\bf 40} (1975), 781--786.

\bibitem{LepLev} F.~Leppington and H.~Levine, On the problem of closely separated circular discs at equal potential,
Quarterly J.\ Mech.\ Appl.\ Math.\ \textbf{25} (1972), 225--245.

\bibitem{LL} E.~H.~Lieb and W.~Liniger, Exact analysis of an interacting Bose gas. I. The general solution and ground state energy,
Phys.\ Rev.\ \textbf{130}(4) (1963), 1605--1616.

\bibitem{love} E.~R.~Love, The electrostatic field of two equal circular conducting discs, Quart.\ J.\ Mech.\ Appl.\ Math.\
\textbf{2} (1949), 428--451.

\bibitem{mcg} J.~B.~McGuire, Study of exactly soluble one-dimensional $N$-body problems,
J.\ Math.\ Phys.\ \textbf{5} (1964), 622--636.

\bibitem{popov} V.~N.~Popov, The theory of one-dimensional Bose gas with point interaction, Theor.\ Math.\ Phys.\ \textbf{30}
(1977), 222--226.

\bibitem{sneddon} I.~N.~Sneddon, \textit{Mixed Boundary Value Problems in Potential Theory}, North-Holland Pub.\ Co., 1966.

\bibitem{S1} B.~Sutherland, Further results for the many-body problem in one dimension,
Phys.\ Rev.\ Lett.\ \textbf{20} (1968), 98--100.

\bibitem{S2} B.~Sutherland, \textit{Beautiful Models: 70 Years of Exactly Solved Quantum Many-Body Problems}, World Scientific
Pub.\ Co., 2004.

\bibitem{Tak} M.~Takahashi, Ground state energy of the one-dimensional electron system with short-range interaction. I, Prog.\ Theor.\ Phys.\ \textbf{44} (1970), 348--358.

\bibitem{TW} C.~A.~Tracy and H.~Widom, On the ground state energy of the delta-function Bose gas, J. Phys. A: Math. Theor. \textbf{49} (2016), 29400.

\bibitem{w} H.~Widom, Toeplitz determinants with singular generating functions, Amer. J. Math. {\bf 95} (1973) 333--383.

\bibitem{ww} E.~T.~Whittaker and G. N. Watson, \textit{A Course of Modern Analysis}, Cambridge Univ.\ Press, 1952.

\bibitem{yang} C.~N.~Yang, Some exact results for the many-body problem in one dimension with repulsive delta-function interaction,
Phys.\ Rev.\ Lett.\ \textbf{19} (1967), 1312--1315.

\end{thebibliography}
\end{document}